# Architecting Non-Volatile Main Memory to Guard Against Persistence-based Attacks

Fan Yao
University of Central Florida
fan.yao@ucf.edu

Guru Venkataramani
George Washington University
guruv@gwu.edu

## ABSTRACT

DRAM-based main memory and its associated components increasingly account for a significant portion of application performance bottlenecks and power budget demands inside the computing ecosystem. To alleviate the problems of storage density and power constraints associated with DRAM, system architects are investigating alternative non-volatile memory technologies such as Phase Change Memory (PCM) to either replace or be used alongside DRAM memory. While such alternative memory types offer many promises to overcome the DRAM-related issues, they present a significant security threat to the users due to persistence of memory data even after power down.

In this paper, we investigate smart mechanisms to obscure the data left in non-volatile memory after power down. In particular, we analyze the effect of using a single encryption algorithm versus differentiated encryption based on the security needs of the application phases. We also explore the effect of encryption on a hybrid main memory that has a DRAM buffer cache plus PCM main memory. Our mechanism takes into account the limited write endurance problem associated with several non-volatile memory technologies including PCM, and avoids any additional writes beyond those originally issued by the applications. We evaluate using Gem5 simulator and SPEC 2006 applications, and show the performance and power overheads of our proposed design.

## 1. INTRODUCTION

With the growing prevalence of multi-core and accelerator-rich processor architectures, the DRAM main memory subsystem is facing increasing pressure to retain the working sets of all of the threads executing on the individual cores. This is necessary to guarantee high application performance and avoid expensive trips to disks during program execution, as well as sustain scalable performance under limited energy and power budgets. Consequently, system architects have begun to look for alternative memory types such as Phase Change Memory (PCM) with higher storage density and comparable performance that show viable promise as a substitute for DRAM technology [1, 2].

Non-Volatile Memory (NVM) technologies, such as PCM, can offer power benefits by removing the need for periodic DRAM refreshes necessary to retain the memory contents. Also, when the system enters one of the power-saving modes (low-power or sleep states), NVM retains the program code and data due to their non-volatile nature. Consequently, when the system enters the active state again, the application can resume without any significant performance penalty that may otherwise occur in DRAM owing to its inability to retain the memory contents without the refresh operations in certain system low-power states.

Despite such powerful advantages, the NVM can be a significant security challenge for user data privacy, and may potentially lead to illegitimate data exfiltration to unauthorized users especially when the runtime system is not actively protecting its data (e.g., when the system is entirely powered down or when the runtime is suspended due to the system being in a low power mode). A malicious user who has physical access to the system could simply extract the sensitive secrets that may be stored inside the memory (say, after system power down) and/or overwrite the memory contents to hijack the program execution (say, after the system resumes to active state). Such scenarios can lead to severely compromising the user's privacy or have detrimental effect on the execution of some important user applications. Therefore, it is necessary to investigate methods that improve the security of NVM and defend them against persistence-based memory attacks.

We note that a key consideration for a solution proposal that seeks to defend many NVM types (including PCM) against memory persistence-based threats is to be cognizant of their limited write endurance issue. That is, NVM technologies such as PCM are expected to sustain an average of $10^8$ to $10^{10}$ writes per cell throughout their lifetime, after which the cell's programming element breaks and the write operations can no longer change the values of the memory cells [3]. Therefore, it becomes necessary to reduce or avoid additional memory write operations beyond those that are issued by the system normally. Even if a future NVM technology were to support unlimited write endurance (e.g., spintronic memory technologies such as MRAM), we note that avoiding additional, unnecessary memory write operations will immensely benefit the overall system performance by reducing memory bandwidth traffic. Therefore, our design consideration to avoid (or at least, significantly reduce) additional memory writes will be the key to enable smooth integration of security-enhancing features of NVM types into the future system architectures.

In this paper, we present an NVRAM data protection scheme that encrypts the data before being written to NVM and performs decryption upon data read using a randomly generated key-pair that is unique for every session, where a session is defined as the interval between two consecutive system power-down events. During a session, the system may be in one of the active or low-power states. To minimize the overheads of encryption/decryption and reduce the impact on memory bandwidth, we explore the use of hardware optimizations such as DRAM buffer cache and evaluate our design. Our exploration offers valuable insights into how the user data can be protected with minimal effect on main memory and overall system performance.

In summary, the contributions of our work are:

1. We investigate mechanisms to defend non-volatile main memory against persistence-based memory attacks by encrypting the data prior to memory writes. Our proposed technique takes into account the limited write endurance problem of many NVM technologies, and avoids exacerbating the problem through encrypting the incoming data prior to memory residency.

2. We show the effect of using various encryption algorithms and explore the use of a probabilistic, differentiated encryption (i.e., the use of different encryption techniques) based on the security needs of the application at various phases. We examine the use of OS/runtime support to identify security critical pages, and outline methods to minimize the impact on application performance.

3. We analyze the overheads of encryption on performance-friendly optimizations such as the use of DRAM buffer cache for NVM main memory.

4. We describe our design and implementation, and demonstrate our experimental findings using a range of applications from SPEC 2006 benchmark suite [5] with varying levels of memory activity. Our results show that our proposed techniques hold good promise to provide defenses against data persistence based attacks while incurring low performance overheads (about 8.8% worst case performance overhead and 3.8% worst case power overhead in memory-intensive benchmarks).

The rest of the paper is organized as follows: Section 2 describes our threat model and assumptions, Section 3 presents an overview of our design, Section 4 has details on our implementation and optimization strategies, Section 5 presents our evaluation setup and experimental results, Section 6 contains related work, and Section 7 presents our conclusions.

## 2. THREAT MODEL

In this work, the fundamental threat that we seek to mitigate is the attacker having physical access to the main memory subsystem and his ability to exploit the persistence property of NVM to read remanent sensitive data and/or manipulate memory contents that may lead to erroneous, harmful program behavior.

In general, there are two possible ways that an attacker could seek access to the NVM contents – 1. An online attack is one where a malicious application can read the working set memory of an application that ran previously prior to the system power down. 2. An offline attack is one where an application can probe the memory after the system is powered down [4]. Note that most traditional OS security models assume that the main memory contents do not persist between system reboots, and hence do not have any explicit mechanisms to protect memory contents after system is powered down.

Also, after exiting the system sleep (low-power) modes, most NVM technologies (such as MRAM [6]) offer instant power-on benefits with readily available memory contents for use due to their non-volatile nature. This is because NVM does not depend on memory refresh operations to retain their contents as seen in DRAM. This normally provides the applications with an added performance benefit of not having to re-fetch memory contents that are potentially lost without the DRAM refresh operation during system sleep. However, an adversary could exploit this NVM property for malicious purposes. That is, when the runtime system (which is responsible for permission and security checks) is suspended during system sleep, a malicious user may modify the NVM contents to cause harmful program behavior. To counter such threats, it is necessary to make sure that the attacker does not gain any meaningful data while the runtime system is not actively operating on (and protecting) the NVM.

In this work, our aim is to address data remanence based attacks in NVM, and *not to address any and all* of the data privacy threats that may occur while the system is active. We assume that the runtime system is capable of actively protecting the memory contents through traditionally well-known memory protection and isolation mechanisms. A computer system, that already has a compromised runtime system or OS, has much greater challenges to address than having to be concerned about memory persistence property of NVM. Therefore, we view the attacks involving compromised applications, runtime or OS to be beyond the scope of our work.

Further, we note that memory-related attacks while the system is in operation are not unique to NVM, and can occur in any system including DRAM, e.g., data thefts that involve snooping of the processor-memory bus, tampered IP blocks, covertly communicating processes that exploit memory as storage channels etc. Any existing technique that protects the memory contents and preserves data privacy during system runtime should be applicable to NVM as well. Our primary goal here is to seek a better understanding of how incorporation of NVM in future systems presents *newer* security challenges for the system architectures to consider that are *normally not seen* in present computer architectures. As such, memory persistence property of NVM, that is absent in traditional DRAM systems, can present newer class of challenges to the OS and hardware architects in terms of providing memory protection and data privacy for the system users.

## 3. DESIGN OVERVIEW

In this section, we present an overview of our technique to counter the data persistence-based attacks on NVM, and outline our approach to hardware design that achieves the desired solution.

In order to address our threat model in Section 2, the main memory subsystem that uses NVM technology should not reveal the memory contents during the system power down that are normally assumed to be lost due to the volatile nature of present DRAM-based systems. To prevent the malicious users from accessing potentially sensitive main memory data, we use hardware encryption modules (that use encryption algorithms like DES, AES, RSA) to encrypt the data prior to memory writes.

Upon every system reboot, a brand new, random encryption key is generated to encrypt the main memory traffic. This is acceptable because the main memory subsystem is not expected to (should not) retain data between system restart operations. This new key generation directly helps to avoid online attacks where a malicious application reads the data left in the NVM prior to restart (see Section 2). The encryption keys are stored in special hardware registers that are accessible only in the privileged mode. When the system enters a low-power (sleep) state where the NVM is gated, the OS clears the hardware registers, and stores the encryption keys as part of its kernel state that can be resumed after the system is active again. This allows the NVM to protect its contents from persistence based attacks while maintaining its performance edge over DRAM-based main memory because of data non-volatility even during system sleep periods. Lastly, since the encryption keys are privileged data only accessible to a trusted OS, an abrupt system restart will not expose this sensitive information to a malicious application. If the attacker had the capability to subvert OS protection and performs privilege escalation, we note that the system has been already compromised, and is under much greater trouble compared to the risks posed by memory persistence based attacks.

The choice of the appropriate encryption algorithm can depend on the user preference for data privacy, as well as the security demands versus memory bandwidth tradeoffs of the application under consideration. For instance, while DES encryption technology uses symmetric keys with 56-bit keys that can be easily vulnerable for attacks, fast hardware implementation of DES encryption using Virtex FPGA achieve speeds up to 10.7 Gbit/sec under maximum clock frequency of 168 MHz [5]. AES, on the other hand, uses substitution-permutation network for encrypting data and is generally more secure than DES. It has reported throughput of up to 121 Mbit/sec at operating clock frequency of 153 MHz and an optimized implementation consumes 3.1k gates [6]. Optimized hardware implementations of RSA cryptography algorithm, which utilize asymmetric keys to enhance security, can achieve maximum throughput of about 1 Mbit/sec for RSA 521b keys and consume 100k gates [7]. Given the differences in speed and security offered by these various hardware implementations of encryption algorithms, we let the user to select the hardware encryption of her choice based on her needs. For this purpose, the memory controller is provisioned with programmable gates to enable the adoption of the chosen encryption algorithm.

Prior to actual main memory writes, the data corresponding to memory addresses are encrypted. Unlike SRAM caches that store data in 32 or 64 byte blocks, the main memory stores the data in larger granularity pages (typically, 4 KB). Typically, the main memory row buffer holds the currently open page, and acts as a fast access cache for the page containing the requested memory address. If the data being written is already present in the row buffer, encryption can be delayed until the memory page is replaced and written back into NVM. In case of row buffer misses, the encryption algorithm needs to decrypt the page first and then perform the write operation. If the successive writes are to different main memory pages, the overhead of decryption/encryption may begin to become noticeable. Also, oft-accessed memory pages may suffer from having to be decrypted frequently. To avoid repetitive de(en)cryption latencies on the same page for successive writes, an efficient strategy is to utilize a write buffer that can combine multiple write operations associated with a single page [8]. This also helps avoid repetitive invocation of the encryption/decryption algorithm, and alleviates the associated performance and power overheads that may otherwise become significant.

For read operations from the main memory that ought to decrypt the data prior to forwarding the resolved memory request to upper-level SRAM caches, we note that it is sufficient to perform the decryption operation just once when the main memory row buffer is filled. When new data is fetched into the row buffer upon replacement, the current row buffer contents are re-encrypted and written back to the NVM.

We note that it is crucial to perform encryption *prior to data writes to the NVM* due to two important reasons:

1. When the system is not properly powered down (through an OS shutdown), the data stored in plaintext format on the non-volatile main memory could be exposed to the malicious hackers, leading to persistence-based memory attacks.

2. Performing encryption on data already resident in NVM increases the total number of writes to the device. This is detrimental to the NVM (such as PCM) that already suffer from limited write endurance problem. Therefore, it is essential to perform encryption before writes to main memory in order to avoid accelerating the NVM device failure.

Figure 1 presents an overview of our design approach. The processor cores issue the data read and write operations to the on-chip SRAM caches. During a last-level cache (LLC) miss, the data requests are sent to the main memory where an encryption/decryption operations are performed on the memory-related traffic.

In order to minimize frequent writes to NVM and to avoid performance loss for such memory-bound applications, a hybrid PCM-DRAM memory configuration with a relatively smaller DRAM buffer that acts as a faster-access cache to retain the working set of the frequently accessed main memory pages can be used [1]. However, for applications that can fit their working set into the caches most of the time and incur fewer trips to the main memory in the first place, the DRAM buffer cache may unnecessarily increase power consumption. We envision the DRAM buffer cache as a feature that may be optionally enabled when a memory-bound application is running. On occasions where the user observes the DRAM buffer to be less utilized (e.g., data access are always SRAM cache hits or DRAM buffer reuse rates are extremely low), the DRAM buffer cache can be disabled and the last-level cache misses can bypass it to access the non-volatile main memory directly. The memory controller hardware that issues memory requests is responsible for enabling and disabling the DRAM buffer. For this purpose, the memory controller is provisioned with a programmable function that turns on the DRAM buffer upon request. This on-demand DRAM availability can be controlled by the OS based on its knowledge of memory activity of the applications running in the system. We note that many modern processors support hardware counters to measure the number of main memory reads/writes and the associated stall cycles (performance penalty), and make them available to the users via special ISA support and software interface [9]. Using such mechanisms, the OS can request to turn on DRAM Buffer cache to minimize the performance impact of data encryption on NVM writes during application execution.

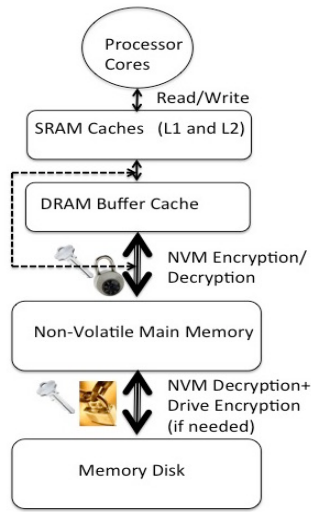

Figure 1: Overview of our Self-obscuring NVM Approach

Finally, we note that there is an increasing demand for Self-encrypting Drive (SED) solutions based on Trusted Computing Group specifications to enable integrated encryption and access control within the protected hardware of the hard drive [10]. TCG's open standards have allowed for multi-vendor interoperability, and let the application developers to operate with multiple SED providers. SED specifies the standards for data confidentiality, and offers hardware-based encryption solutions within the hard drive electronic circuits. We note that such an openly available standard for hard drive encryption offers us two possibilities in interfacing with non-volatile main memory:

1. We could investigate mechanisms to integrate our encryption based solution for NVMs with the already available hardware-based SED solutions. Essentially, this solution calls for extension of the SED technology to NV main memory. Though this is useful, we note that the main memory is not expected to retain data between system-reboots unlike the memory disk that is expected to retain (and preserve) the user data throughout its lifetime. Therefore, the constraints posed by data privacy needs of NVM-based memory might be less stringent compared to disks.

2. We could decrypt the pages replaced from NVM, and allow the SEDs to re-encrypt the data using their encryption technology before being written to disks. Since the disk-main memory data traffic is usually off-the-performance-critical path for most applications (with a few exceptions like data-streaming media applications), this option may be more acceptable and cheaper for a vast majority of cases. In our current design, we incorporate this solution to handle SED-based memory disks (see Figure 1).

## 4. IMPLEMENTATION

Popular encryption algorithms such as DES, AES and RSA have been widely studied, and a variety of optimized implementations exist in hardware. Many existing encryption algorithms are typically faster than the actual main memory accesses, effectively hiding the encryption latency within the actual memory access times. For instance, depending on the processor frequency, optimized AES implementations consume anywhere between 25-40 ns in time [11]. However, access time for a 64-byte cache line, based on a 4 ns memory bus clock cycle, is expected to be 140 ns [4]. If the data hits in the row buffer, the memory access latency is even lower. Note that the row buffer data is already decrypted upon memory read, and does not need encryption until data is written back to the NVM. To further keep the performance impact minimal on certain memory-bound applications, hardware optimizations such as DRAM buffer cache can be useful because they maintain the application's current working set and act as a fast access cache for NVM.

Even when the performance impact can effectively be hidden through optimizations, frequent invocation of encryption could adversely impact other system parameters such as power and energy consumption. To reduce runtime impact on application power (and sometimes performance too), a first-order solution can be to use probabilistic and differentiated encryption where individual memory banks are encrypted using different encryption algorithms. Naively speaking, the type of encryption used within a particular bank could be randomly chosen just to prevent remanent data from being stolen easily, and the encryption keys belonging to the different memory banks are stored separately within the memory controller. This solution offers varying levels of security to the different memory banks depending on the encryption technology used. This differentiated approach leads to reduced power and performance overheads especially when many of the memory banks can afford to use low-cost encryption techniques at the cost of potentially reduced security for non-critical data. An un-informed choice of algorithm in differentiated encryption can lead to potentially compromising sensitive data that may be otherwise unavailable to the adversary when stronger encryption techniques are used.

A more effective alternative is when the OS and the user could provide hints to the memory controller on the sensitivity levels of data (e.g., the memory page contains important kernel data structure pointers, the application has sensitive user data inside the memory page). Based on the level of protection needed, the OS assigns different security flags (distinguishing the potent for threat) to the individual memory pages. Based on the OS-assigned security flags for a memory page and the highest security demand among an individual bank's constituent pages, the memory controller chooses an appropriate encryption algorithm for the entire memory bank. Furthermore, with a simple memory page map (maintained in the OS) to indicate the need to encryption/decryption, the choice to invoke encryption algorithm can be made at the granularity of individual memory pages, thus eliminating the need to encrypt all memory pages within a bank. Note that, the choice of encryption algorithm for an individual bank is still determined by the page that demands the highest security within a bank.

The information needed to decide on whether to encrypt memory pages and the appropriate level of security needed can be derived statically based on the applications profile information from the OS, and can be dynamically re-calibrated depending on the application phases. As an instance, an application could work on sensitive data for a certain timeframe during when high levels of data privacy is needed, and may work on non-sensitive data during other periods of time when stringent security can be relaxed and traded with less secure encryption. If the application's working data sets do not overlap between two such distinct phases, the memory bank can switch to a lower-cost encryption algorithm. To avoid any unintended, yet benign, reads from the now-garbage memory pages (rendered unreadable because of a different encryption algorithm during a previous application phase), the OS can choose to explicitly mark these pages as invalid. If the working set pages between two different application phases are non-overlapping, this explicit marking of invalid pages by the OS should not incur any performance overheads. In the event that the working set memory pages between two application phases are indeed overlapping, there are two alternatives: 1. Keep using the encryption algorithm that was used in the previous phase and pay additional costs that may be now-unnecessary for the reduced level of security needed, 2. Switch to a lower cost encryption, and invalidate the now-garbage memory pages that will incur extra overheads when such pages are accessed across the two phases. The choice of an appropriate alternative depends on the length (execution time) of each application phase and the degree of memory page content sharing between the two application runtime phases with distinct security needs.

In this design space where the OS or application may make the appropriate data sensitivity information available to the hardware, a third aggressive solution strategy would be to completely eliminate encryption on certain memory banks or the individual memory pages based on the application profile information from the OS. Again, we note that the applications could have phase-based behavior with differing security needs. Therefore, it is important to consider the length of application phases and the degree of data sharing between phases to study the usefulness of this differentiated encryption with no encryption during certain phases.

## 5. EVALUATION

In this section, we perform evaluation of our proposed solution strategy that counters persistence-based memory attacks in NVM. We first outline our evaluation setup, and then present our experimental results.

### 5.1 Experimental Setup

Our evaluation platform uses Gem5 [12], a full system cycle-accurate simulator for modeling system-level and processor microarchitectures. We integrated NVMain [13], a cycle-accurate main memory simulator that models emerging non-volatile memory in Gem5. Our baseline models a processor running at 1 GHz, with a private 2-way set associative L1 cache and a shared 2MB, 8-way set-associative LLC. The block size is 64 Bytes in all caches. For the PCM-based NVRAM, we model a single channel, 4-bank, 4 GB capacity with 4 KB pages with read latency of 50 ns and write latency of 1 $\mu$s [3]. For the hybrid memory with DRAM buffer cache, we create a NVMain configuration that features a 128MB DRAM cache with 8 banks that acts as a fast access cache to store the pre-decrypted application's working set and hides the decryption latency incurred for accessing the PCM directly. The DRAM cache is set to share the same memory channel as the off-chip PCM memory. We have incorporated McPAT [14] into our simulation infrastructure by mapping Gem5's simulation output to McPAT compatible data format for processor power modeling. A baseline processor based template file is created for McPAT's internal power calculation. We collected memory power consumption data directly from NVMain's simulation output statistics. We use SPEC 2006 applications [15] with reference inputs, and the applications were run in system call emulation mode. We fast-forward 700 million instructions to skip the initialization phase, and then simulate 500 million instructions in detail.

In our evaluation, we use four different encryption algorithms (DES, AES, RSA and differentiated encryption) under two different memory configurations (PCM NVRAM and hybrid PCM-DRAM) in our evaluation. The hardware implementation of encryption techniques were based on optimized version described in various prior studies [4, 5, 6, 7, 11]. Based on the currently available data in the literature, DES hardware encryption incurs an average overhead of about 7-10 processor cycles per word, AES hardware encryption consumes about 12-15 cycles per word, and the RSA hardware incurs penalty of 24-30 cycles per word. Using this data, we model the encryption overheads as the average number of cycles it takes to encrypt a main memory page for a given encryption algorithm. The Gem5 configuration files are modified accordingly to reflect the corresponding encryption/decryption overheads for different algorithms.

### 5.2 Experimental Results

We ran over 20 SPEC applications. We observed that several non-memory-intensive applications (e.g, GemsFDTD) did not exhibit significant memory activity, and as a result, the execution time (performance) and average power overheads in these benchmarks were negligible relative to a baseline with no encryption. Therefore, in our experimental results section, we show the performance and average power overhead results for six representative benchmarks from our evaluation– *bzip2*, *lbm*, *mcf*, *omnetpp*, *soplex* and *milc*. *mcf* and *lbm* are high memory-intensive applications (high number of LLC misses). The other four applications also have memory-intensive activity from low to moderate levels.

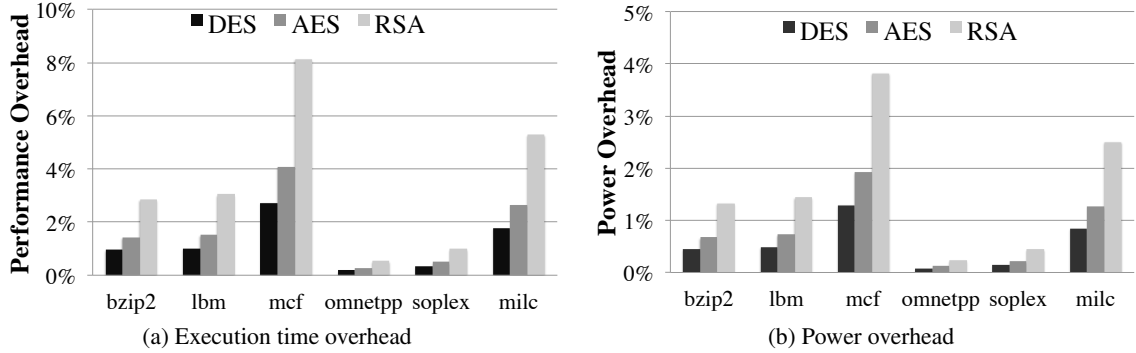

Figure 2: Execution time (performance) and Average Power overheads on SPEC 2006 benchmarks due to different hardware encryption algorithms on 4GB PCM memory. Baseline has PCM NVM with no encryption.

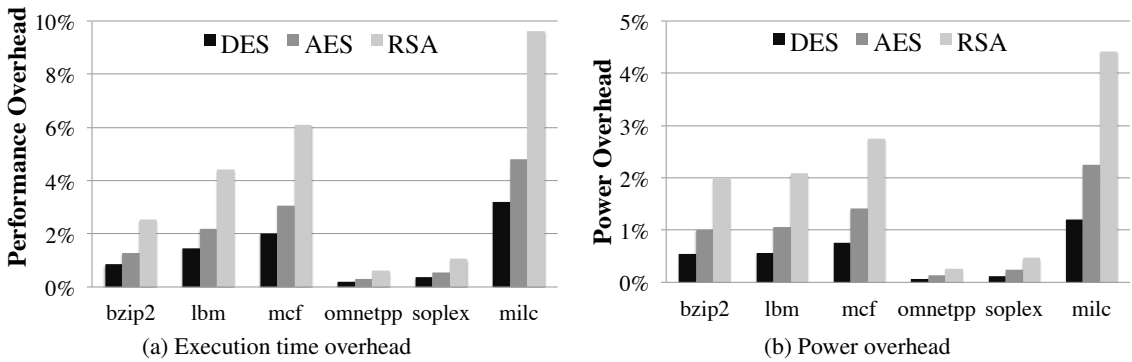

Figure 3: Execution time (performance) and Average Power overheads on SPEC 2006 benchmarks due to different hardware encryption algorithms on 4GB PCM hybrid memory with a 128MB DRAM Cache. Baseline has PCM-DRAM with no encryption.

Figure 2 shows the execution time and power overheads on these six benchmarks using the different encryption algorithms, where the baseline does not have any encryption. We see that the mcf benchmark, with a high degree of memory activity, consistently has higher performance overhead ranging from 2.7% (DES) to 8.1% (RSA), and the corresponding power overhead ranging from 1.3% to 3.8%.

Figure 3 illustrates the performance and average power overheads on a hybrid PCM-DRAM. We observed that by using a 128 MB DRAM buffer cache, the baseline IPC for most memory-bound applications already increased noticeably. This is because the DRAM cache is able to store the active working sets for these applications and thus reduce the number of accesses to the PCM memory. Our experiments show that the DRAM cache had beneficial effect in certain applications by reducing their performance overheads, while having a detrimental performance effect through further widening the execution time gap between the baseline and the encryption in other applications. As examples:

1. In mcf benchmark, when compared to the memory configuration without DRAM cache, the accesses to PCM reduced by almost 63% with a DRAM cache. This resulted in lowering of RSA encryption overhead from 8.1% (without DRAM cache) to about 6% (with DRAM cache). Correspondingly, the power overheads for RSA encryption also reduced from 3.8% (without DRAM cache) to about 2.7% (with DRAM cache).

2. In milc benchmark, when compared to the memory configuration without the DRAM cache, the DRAM cache was only able to filter 15% of the accesses to PCM. With the added DRAM lookup latency in the hybrid configuration and the still-dominant en(de)cryption latencies (due to high DRAM miss rate), the RSA performance overheads increased from about 5.2% (without DRAM cache) to about 8.8% (with DRAM cache).

Based on our discussion in Section 4, we note that the performance and power overheads of encryption can be improved through the use of differentiated encryption strategy that chooses to boost encryption in security-critical phases and adopt low-cost encryption schemes for non-critical phases. To demonstrate the use of our probabilistic, differentiated encryption, we conduct experiments where we assume that the application has highly security-sensitive computations for about 25% of its runtime (RSA is used for encryption), moderately security-sensitive computations for about 60% of its runtime (AES is used for encryption) and less sensitive computation for the remaining time (DES is for protection during this time). This assumption is made to illustrate an average application that has phases of sensitivity during its lifetime. Figure 4 show our experimental results on a PCM based NVRAM where we observe that even memory-intensive mcf exhibit relatively low performance overheads of around 5%. Figure 5 shows the performance and power overheads of our differentiated encryption when a hybrid memory system is used. Our results show that milc benchmark has the maximum performance overhead of 5.9% and power overhead of 2.8% respectively. These results shows that a good security-performance tradeoff can be achieved by customizing the encryption of data based on the security needs of the application phases.

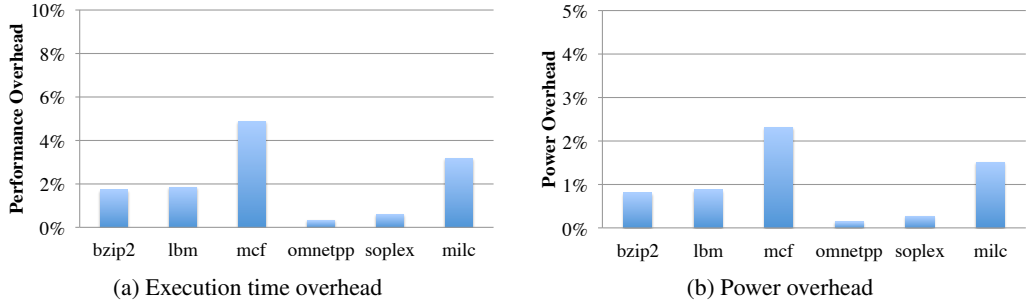

Figure 4: Execution time (performance) and Average Power overheads on SPEC 2006 benchmarks due to different hardware encryption algorithms on a 4GB PCM memory. Baseline has PCM NVM with no encryption.

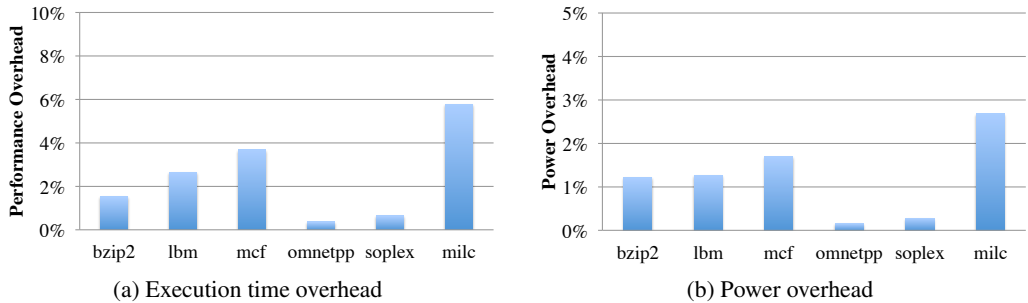

Figure 5: Execution time (performance) and Average Power overheads on SPEC 2006 benchmarks due to different hardware encryption algorithms on a 4GB PCM memory with a 128MB DRAMCache. Baseline has PCM-DRAM with no encryption.

## 6. RELATED WORK

Prior techniques have studied memory persistence based attacks in NVM. In contrast to i-NVMM [16] that incrementally encrypts the memory resident data incurring additional memory writes, our technique performs encryption of data prior to the actual memory writes. In this way, our approach minimizes the exacerbation of limited write endurance problem that have already plagued the widespread adoption of many NVM technologies.

Enck et al. [4] have proposed encryption using a special key stored specially in a smart card, and one-time pads that are added to the data prior to encryption. This method requires extensive modifications to cache blocks such as addition of several state bits to avoid repetition of one-time pad counters. Note that the read operations (that are often performance critical) still suffer due to decryption latency. In contrast, our solution approach explores minimal changes to the memory hardware with features to improve performance overheads via buffer caches and differentiated encryption.

To address the limited write endurance problem and enhance PCM cell lifetime, Qureshi et al. [1] have proposed line-level write back instead of redundant write operations to the entire memory page. We note that such techniques can be utilized in conjunction with wear-leveling techniques [17] to minimize the possibility of early memory cell wear-out in NVM technologies.

Recently Awad et al. [18] observed that the OS does data shredding where newly allocated pages are zeroed out to protect the remnant data from the previous owner process of that page. They noted that data shredding could cause write endurance issues for NVM especially when a large percentage of writes to the Non-volatile memory are for zeroing out the page's contents. To address this, they propose a hardware mechanism to delay the zeroing out the data bit values until when a cache block from a newly allocated page is read by the LLC. This saves unnecessary 'zero' writes to the NVM. We note that this improved data shredding strategy could be combined with our proposed differentiated encryption to enhance both the security and to minimize the performance impact on the application runtime.

Other prior studies [19, 20] have proposed encryption architectures that provide tamper-proof computations and mem-ory protection, and mechanisms to protect virtual memory [21]. Our proposed solution can work synergistically with all of such prior solutions to enhance the NVM security. In addition to improving memory reliability, we note that other mechanisms to improve software robustness [22, 23, 24, 25, 26, 27, 28] will improve overall system security.

## 7. CONCLUSION

In this paper, we presented smart mechanisms to leverage the hardware encryption methods for enhanced NVRAM security against data remanence attacks, and explored techniques to make it harder for an adversary to steal the persistent data left in the non-volatile memory after power down or in-between system sleep periods. Our mechanism takes into account the limited write endurance problem associated with several non-volatile memory technologies including PCM, and avoids any additional writes beyond those issued by the system normally. We showed experimental results demonstrating the performance and power overheads of using encryption techniques on two different memory configurations- a PCM-based NVRAM and a hybrid PCM-DRAM memory configuration. We also studied differentiated encryption strategy where the encryption algorithm is chosen based on the security needs of the application phases. We showed that this strategy delivers security for the critical application phases and improves the performance of other non-security-critical phases of the application. Our experimental results show that our mechanisms incur low performance penalty even in memory-intensive applications and exhibit low average power overheads.